\documentclass[12pt,a4paper]{iopart}

\usepackage{graphicx}

\begin{document}

\title[Scalable quantum computing stabilised by optical tweezers on an ion
crystal]{Scalable quantum computing stabilised by optical tweezers on an ion
crystal}

\author{Yu-Ching Shen and Guin-Dar Lin}

\address{Center for Quantum Science and Engineering and Department of Physics,
National Taiwan University, Taipei 10617, Taiwan}
\ead{guindar.lin@gmail.com}
\vspace{10pt}
\begin{indented}
\item[]October 2019
\end{indented}

\begin{abstract}
As it has been demonstrated that trapped ion systems have unmatched
long-lived quantum-bit (qubit) coherence and can support high-fidelity
quantum manipulations, how to scale up the system size becomes an
inevitable task for practical purposes. In this work, we theoretically
analyse the physical limitation of scalability with a trapped ion
array, and propose a feasible scheme of architecture that in principle
allows an arbitrary number of ion qubits, for which the overhead only
scales linearly with the system size. This scheme relies on the combined
ideas of a trap architecture of tunable size, stabilisation of an
ion crystal by optical tweezers, and continuous sympathetic cooling
without touching the stored information. We demonstrate that illumination
of optical tweezers modifies the motional spectrum by effectively
pinning the ions, lifting the frequencies of the motional ground modes.
By doing so, we make the structure of the array less vulnerable from
thermal excitations, and suppress the the position fluctuations to
insure faithful gate operations. Finally, we also explore the local
behaviour of cooling when a sub-array is isolated by optical tweezers
from other parts of the crystal.
\end{abstract}

%
%
%
%
%

\section{Introduction}
As some prototype quantum computing (QC) resources have
emerged in the commercial market for recent years, there have been
more and more user experiences and exploding interest in quantum technologies
and algorithms that aim to demonstrate its supremacy over any classical
means. In addition to the solid-state realisation based on superconducting
circuits, trapped ion systems remain one of the most promising alternative
platforms. Up to present, they retain the record ($\sim$99.9\%) of
highest fidelities of quantum state and gate manipulation \cite{record_ballance16,record_gaebler16,fastgateexp18},
and the largest size of engineered qubit entanglement \cite{14ghz11,200entanglement16}.
The latest benchmark test report released by IonQ Inc. has shown very
competitive results \cite{ionq_benchmark}. Further, such systems
offer a perfect bottom-up approach to investigate many-body correlation
and quantum simulation \cite{nonlocal_monroe14,quasiparticle_roos14,imaging_spin_monroe14}
as well as other perspectives in the study of atomic clocks, non-classical
motional states, and phonon lasers of growing interest very recently.

A typical ion trap, as known as the Paul trap, is constructed by time-varying
radiofrequency (RF) electric field, which confines and aligns ions
in a linear structure. In QC, each ion serves as a quantum-bit (qubit)
as the information is encoded in two of the atomic states. The qubit
degrees of freedom are protected from ions\textquoteright{} motion
unless the coupling between motion and qubit states are turned on
by applying state-dependent forces. During the quantum gate processing,
the collective motion is used as the quantum bus. The entanglement
between motion and qubit states is crucial for gate operations, but
needs to be removed when the operation is completed. Any residual
entanglement introduces computational errors and heat for subsequent
operations. The most commonly used scheme nowadays is the Mølmer-Sørenson-Milburn
gate \cite{sorensen99,milburn00}, which utilizes bi-chromatic interference
to eliminate temperature dependence so can be operated at higher temperatures.
It is further shown that, by operating the transverse modes \cite{zhu06},
a faithful gate can be accomplished requiring only the Doppler cooling
\cite{lin09}.

Though ion systems excel other platforms in many ways such as long-lived
atomic coherence, perfectly identical qubits, strong laser-mediated
Coulomb interaction, and deterministic control engineered entanglement,
the major challenge is its scalability. There have been several scalable
proposals including ion shuttling \cite{shuttling02,shuttling_blakestad_11,shuttling_moehring11,scaleup_monroe13},
quantum network \cite{scaleup_monroe13,ionphoton_duan04,rmp_duan_monroe10},
and arrays of microtraps \cite{microtrap_cirac,microtraps_fastgates_hope18},
which, however, introduces new issues of hardware fabrication (shuttling,
microtraps), movement control (shuttling), slow processing speed (shuttling),
and probabilistic and lossy quantum interfacing (quantum network).
The 1D geometry of a Paul trapped ion array is the most straightforward,
but seems very difficult to scale up. Nevertheless, it is still worth
investigating the fundamental reasons of such limitation in 1D for
it might provide informative insights in more complicated structures.
Here, we discuss the scalable issues in the following: (1) Architecture
issue \textendash{} Adding more ions requires building a larger trap.
The desired architecture must be scalable with single qubit addressability.
To avoid nearby ion cross-talking, the ions' spacing $d_{0}$ must
be kept a few microns (in this paper, we take $d_{0}=10$ $\mu$m)
to allow a focused Gaussian beam of width approximately half of the
spacing. Further, it is beneficial to have gate controlling parameters
universal along the crystal, without considering the actual location
where gates are performed as long as the distance of target qubits
(gate distance) is fixed. Thus, the ideal geometry must be uniform
\cite{lin09}. (2) Array stability and cooling issue \textendash{}
For a very large linear ion array, the bottom of the trap is uaually
flat in order for the ions distributed uniformly. In this case, the
lowest frequencies of the longitudinal motion vanish. The ground mode,
for instance, corresponds to collective macroscopic translation, and
may cause qubit ions to move off sites and make them hardly addressable.
Also, these motional states heat up easily due to divergent phonon
numbers, thus jeopardizing the stability of the array structure. (3)
Gate design issue \textendash{} In a large array, it fails to realize
a two-qubit gate in the adiabatic regime due to the unresolvable collective
motional spectrum. Fortunately, gate operation is still possible by
taking into account multi-modes that contribute to local motional
degrees of freedom of involved ions. This suggests usage of fast gate
operation such that the local motion will not `spread' out over
the whole system.  Two major such protocols are the push gate scheme
based on pulsed lasers \cite{fastgate_garciaripoll03,fastgate_duan04,fastgate_steane14,fastgate_hope15}
and the pulse shaping scheme based on continuous-wave lasers \cite{zhu06,lin09,cw_garciaripoll05,arbitraryspeed_zhu06,fastgate_cw_palmero17}.
These fast gates have been shown to have very high fidelity but demand
strong laser power.

In this manuscript, we aim at resolving the architecture and cooling
issues by proposing a scalable trapping scheme for a large linear
ion crystal, stabilised by optical tweezers. Optical tweezers are
dipole traps formed by off-resonant Gaussian beams that can be focused
to about a few microns of beam width so are able to illuminate individual
ions. Typically, dipole trapping forces are much weaker than Coulomb
forces between charged particles so may not be very useful for confining
ions. But for a large array, the Coulomb forces from each other are
cancelled in the mechanical equilibrium positions of ions. The relevant
frequency scale is then determined by the residual Coulomb interaction,
i.e., the next order of the Coulomb interaction while the ions are
perturbatively displaced from the equilibrium. This frequency scale
characterising the momentum exchange between adjacent ions is given
by $\omega_{0}\equiv\sqrt{e^{2}/(md_{0}^{3})}$ (discussed below),
where $e$ and $m$ are the charge and mass, respectively, of an ion.
With $\omega_{0}$ typically ranging from a few hundreds of kilohertz
to megahertz, we will show how the application of optical tweezers
of frequency about the same order of magnitude modifies the motional
spectrum to improve stability of the crystal and help with the cooling
problems.

This manuscript is organized as follows: In Sec.\,\ref{sec:Architecture},
we propose a scalable architecture where a large ion crystal can be
constructed. We analyse the associated motional spectrum and discuss
the stabilisation made by introducing the optical tweezers. In Sec.\,\ref{sec:Gate},
we demonstrate high-fidelity gates based on transverse modes of the
uniform array, focusing on the locality and the translation symmetry
of the controlling parameters. We also discuss the major sources of
gate errors due to thermal noises. Sec.\,\ref{sec:Cooling} investigates
the sympathetic cooling performance on the ion crystal, where we show
that the position fluctuations are suppressed in the presence of optical
tweezers under cooling. Further, in Sec.\,\ref{sec:Localtrap} we
look at the local dynamics of a sub-array when it is exclusively cooled
while isolated by optical tweezers, demonstrating a remarkable feature
of locality that is important for parallel manipulation and cooling.
Finally, we conclude this work by Sec. \ref{sec:conclusion}.

\section{Scalable architecture\label{sec:Architecture}}

\begin{figure}[tb]
\includegraphics[width=12cm]{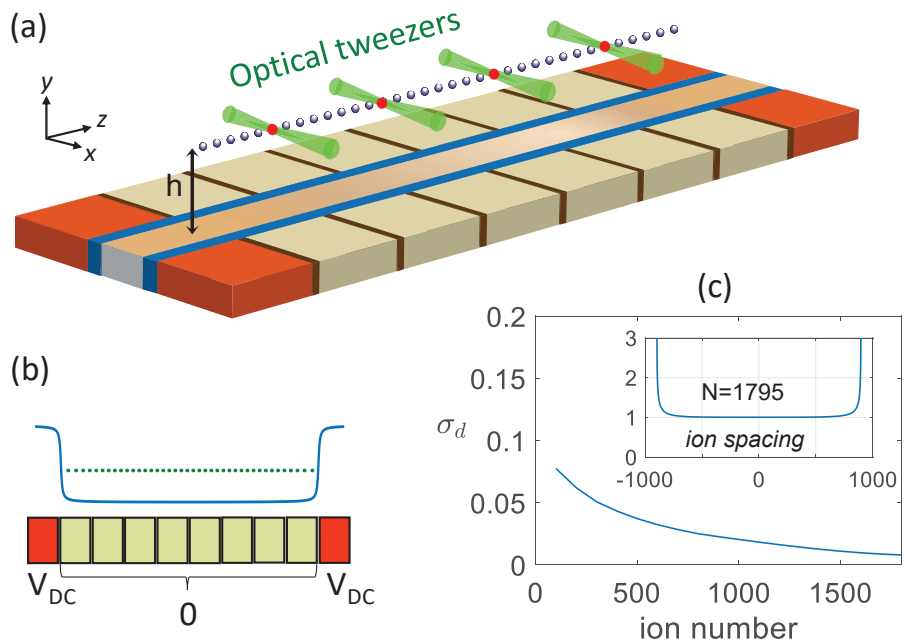} 
\centering{}\caption{(a) Scalable ion trap architecture for a planar realisation \cite{planar_gtri16}
for quantum computing with a series of optical tweezers applied on
regularly distributed ions over the crystal. (b) The surface DC electrodes
build a `bookend'-like potential whose size can be adjusted according
to the system size. (c) Uniformity characterised by standard deviations
$\sigma_{d}$ of ion spacing for various system sizes by keeping the
mean spacing equal to $d_{0}$. Here, we take $L=100qd_{0}$ for $q=1,2,\cdots$,
$20$ as the distance between two bookend walls and $V_{{\rm DC}}=0.1$
V (see text). The mean and standard deviation of spacing are taken
from only the 80\% ions in the middle of the crystal. (Inset) The
spacing distribution for an ion crystal of length $N=1795$ in units
of $d_{0}$. \label{fig:architecture}}
\end{figure}
\noindent A typical Paul trap is shown in Fig.\,\ref{fig:architecture}(a).
Here, we take the planar trap for example \cite{planar_gtri16}, but
our following analysis applies to other realisations. To contain a
uniform large 1D array, the trapping potential must have a flat bottom
with both ends terminated by two walls, resembling a pair of `bookends'
as depicted by Fig.\,\ref{fig:architecture}(b). The trap size (separation
between two bookends) should be adjustable in order for holding more
ions. Fig.\,\ref{fig:architecture}(a) demonstrates a practical implementation,
where two parallel RF wires (in blue) provide the transverse confinement
that supports the ion array above by a separation $h$; along the
axial direction paves a segmental structure of electrodes with each
piece separated by thin isolating layers such that each piece can
be applied a different DC voltage. Most of the pieces are set to the
ground except by the trap edges the potential is lifted for global
confinement. Here we describe the bookend potential landscape in Fig.\,\ref{fig:architecture}(b)
by the simplified model:
\begin{equation}
U(z) =\cases{\frac{eV_{\rm DC}}{\pi}
\left(\tan^{-1}\frac{h}{z+\frac{L}{2}}-\tan^{-1}\frac{h}{z-\frac{L}{2}}\right)
&for $|z|\le\frac{L}{2}$\\
\frac{eV_{{\rm DC}}}{\pi}
\left(\tan^{-1}\frac{h}{z+\frac{L}{2}}-\tan^{-1}\frac{h}{z-\frac{L}{2}}+\pi\right)
&for $|z|>\frac{L}{2}$}\label{eq:longpotential}
\end{equation}

\noindent where the voltage applied to the edge electrodes is $V_{{\rm DC}}>0$,
and the bookend-to-bookend distance is $L$. By taking $h=30$ $\mu$m
and $V_{{\rm DC}}=0.1$ V, we calculate the classical equilibrium
positions of the ions. When the ion number gets large ($>10^{2}$),
the array can be made rather equally spaced if we only take the middle
80\% of the array into account. The uniformity characterised by the
standard deviation is shown in Fig.\,\ref{fig:architecture}(c),
where we observe that the uniformity is less than 4\% for $N>500$.
The actual spacing distribution is also plotted for $N=1795$ and
$L=2000d_{0}$ with $d_{0}=10$ $\mu$m. Note that the unequal spacing
near the edges is inevitable since those ions experience non-uniform
forces from rest of the array and the bookend potential. But they
can be excluded from computing tasks due to lack of homogeneity.

As any part of an array might expose to environmental noises and collisions,
a larger system is more vulnerable for being more likely to be disturbed.
These disturbances contribute to heating in a sense, and can be dealt
with by cooling techniques. The associated fundamental difficulty
is that the lowest motional frequency $\omega_{{\rm L}}$ vanishes
as $N\rightarrow\infty$, so the position fluctuations and the phonon
number $\bar{n}\sim k_{{\rm B}}T/(\hbar\omega_{{\rm L}})$ diverge
for any given finite temperature. To circumvent this problem, we propose
to apply optical tweezers to alter the motional spectrum. Here, a
characteristic frequency scale is determined by the next order of
the Coulomb interaction, that is, the residual term $e^{2}/d_{0}^{3}\equiv m\omega_{0}^{2}$,
which gives $\omega_{0}\equiv\sqrt{e^{2}/(md_{0}^{3})}$ that accounts
for the momentum exchange between adjacent ions. Typically, $\omega_{0}$
is about $0.1\sim1$ MHz, such conditions allow dipole traps such
as optical tweezers of frequency on the same order of magnitude to
make significant differences. For an ion array sitting in the equilibrium,
the interaction can be expressed under the harmonic approximation
as $\sum_{i,j}A_{ij}^{\xi}x_{i}^{\xi}x_{j}^{\xi}$ with the coupling
matrix elements $A_{ii}^{\xi}=\nu_{\xi,i}^{2}+\nu_{\xi,i}^{{\rm ot}2}+\sum_{l=1,l\neq i}^{N}c_{\xi}/|u_{i}-u_{l}|^{3}$
and $A_{ij}^{\xi}=-c_{\xi}/|u_{i}-u_{j}|^{3}$ for $i\neq j$, where
$\xi=x$, $y$ denote two transverse directions, and $\xi=z$ the
longitudinal direction; $c_{x}=c_{y}=-1$, and $c_{z}=2$; $u_{i}$
is the equilibrium $z$ position of the $i$th ion in units of $d_{0}$.
Also, $\nu_{\xi,i}=\omega_{\xi,i}/\omega_{0}$ and $\nu_{\xi,i}^{{\rm ot}}=\omega_{\xi,i}^{{\rm ot}}/\omega_{0}$
are dimensionless frequencies. For $^{171}\rm{Yb}^{+}$ ions, we
have $\omega_{0}=2\pi\times143$ kHz for $d_{0}=10$ $\mu$m, which
is taken to be the exemplary system throughout this manuscript. In
addition, the RF transverse confinement frequency $\omega_{x,i}=\omega_{y,i}=2\pi\times5$
MHz is the same for all ions. The longitudinal one is given by the
second-order derivative of the potential (\ref{eq:longpotential})
$\nu_{z,i}^{2}(z)=(\omega_{0}^{2}m)^{-1}\partial^{2}U/\partial z^{2}$
evaluated at $z=u_{i}d_{0}$. The frequency $\omega_{\xi,i}^{{\rm ot}}$
is provided by the optical tweezers experienced by the $i$th ion
($\omega_{\xi,i}^{{\rm ot}}=0$ means no tweezers). Here we also assume
that the optical tweezer beams, if present, are incident along the
$x$ direction so that $\omega_{y,i}^{{\rm ot}}=\omega_{z,i}^{{\rm ot}}>0$.
Along the $x$ direction of incidence, even though there is a trapping
effect due to spatial variation of a focused Gaussian beam, we take
$\omega_{x,i}^{{\rm ot}}\approx0$ because it is usually much smaller
than the RF one $\omega_{x,i}$.

\begin{figure}[tb]
\begin{raggedleft}
\includegraphics[width=12cm]{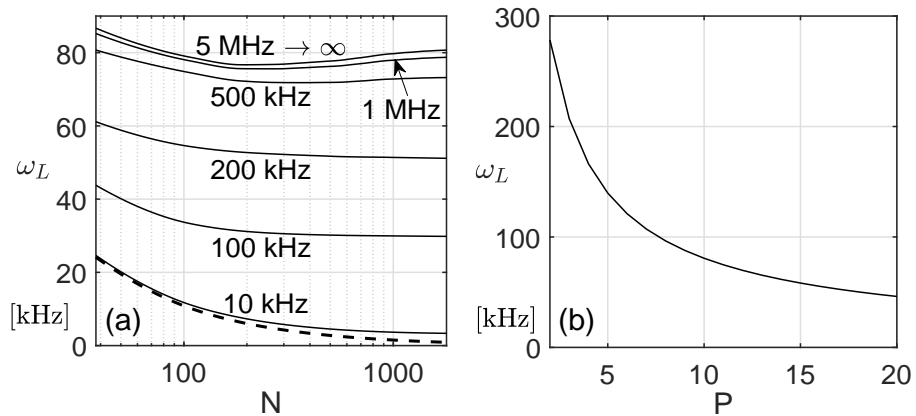} 
\par\end{raggedleft}
\centering{}\caption{(a) The lowest longitudinal frequency in kHz with various optical
tweezer strength for different system sizes $N$. The optical tweezers
are applied every $P=10$ ion sites. For comparison, the dashed curve
represents to tweezer-free cases. Note that for the tweezer strength
larger than $2\pi\times1$ MHz, the systems start to display similar
motional spectral distribution. This means that the `tweezered ions'
can be effectively treated completely frozen. (b) Lowest frequency
in kHz as a function of repetition period $P$ for $N=1795$ and $L=2000d_{0}$.
In all cases, we take  $V_{{\rm DC}}=0.1$ V and $h=30$ $\mu$m.
\label{fig:tweezerf}}
\end{figure}

To see how optical tweezers help stabilise a large ion crystal, we
first consider the arrangement shown in Fig.\,\ref{fig:architecture}(a).
Here, a series of optical tweezers are regularly distributed shining
on ions with a spatial repetition period $P=10$ sites over the system.
Hereafter, the ions that are illuminated by optical tweezers are called
`tweezered' ions in this manuscript. The application of optical
tweezers amounts to introducing separators partitioning the whole
ion crystal into `cells'. An array of cells of the same size is
thus constructed by inserting these optical tweezers with each cell
containing $P-1$ ions. In Fig.\,\ref{fig:tweezerf}(a), we show
that, for $P=10$ and $N\sim10^{3}$, the lowest longitudinal frequency
$\omega_{{\rm L}}$ can be raised by one to two orders of magnitude
by simply applying tweezers of frequency up to 500 kHz. With tweezer strength
larger than $5$ MHz, the frequency profile appears to converge to the
case of infinite tweezer strength, meaning that these tweezered ions
are effectively pinned in space.  Fig.\,\ref{fig:architecture}(b)
shows the lowest frequency against the period $P$ of the tweezer
distribution, where we find $\omega_{{\rm L}}/\omega_{0}\approx3.36P^{-0.77}$.
This frequency scale is comparable to that of a single cell, implying
that the corresponding modes are determined dominantly by local degrees.
The emerging localization of motion is a remarkable feature as the
tweezer frequency becomes sufficiently strong, or multiple tweezered
ions are bundled to be treated as a separator. This point will be
discussed in the context of cooling in Sec.\,\ref{sec:Localtrap}.
On the other hand, since optical tweezers are off-resonant dipole
traps, in principle they do not evoke population transitions. But
AC Stark dephasing may still occur and degrade the coherence. Though
it can be somewhat calibrated, it should be avoided to store quantum
information in tweezered ions. Fortunately, we hold the flexibility
to turn optical tweezers on and off at any time. That is, tweezered
ions are recyclable.

\section{\textit{Quantum gate and fidelity\label{sec:Gate}}}

\noindent Quantum gates are implemented by coupling internal and motional
states, exchanging quantum information via common collective phonon
modes. On a large ion array, the motional spectrum becomes rather
complicated and hardly resolvable. Therefore, a working quantum gate
needs to involve multiple motional modes at the same time. Superposition
of these modes in fact contributes to local motion of the target ions
and their neighbouring ones. Given that the ion crystal is uniform,
the local degrees associated with such a quantum gate must have translation
symmetry, i.e., the control parameters are universal no matter where
the gate is implemented as long as the distance of the two target
qubits is given. Quantum gates over larger distance demand stronger
laser power. Finding the optimized control parameters over laser power
or gate times is however not the scope of this manuscript. We here
focus on the effects on the gate fidelity when the system scales up
according to the proposed scheme stabilised by optical tweezers.

\begin{figure}[t]
\begin{raggedleft}
\includegraphics[width=15cm]{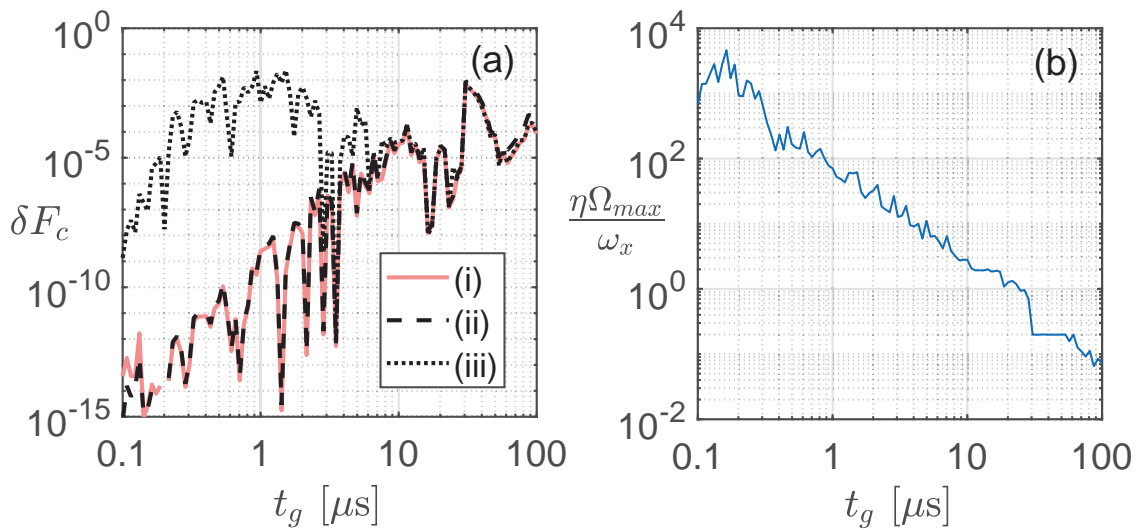} 
\par\end{raggedleft}
\centering{}\caption{(a) The computational infidelities of a transverse-mode (the $x$
or $y$ directions) CPF gate using $M=7$ segment pulses for different
gate times, using the pulse shapes obtained by considering a small
array of $N=9$ and optimizing the fidelity over choices of $\mu$
for a given $t_{g}$ and $T=10T_{{\rm D}}$ with $T_{{\rm D}}$ the
Doppler temperature. Curve (i) corresponds to the original infidelity
with the $3$rd and $6$th ions on the $N=9$ array. Curve (ii) corresponds
to the infidelity of the transverse $x$-mode gate between the $628$th
and $631$st ions (chosen randomly) on an $N=1795$ ion crystal. Curve
(iii) corresponds to that of the transverse $y$-mode gate between
the $624$th and $627$nd ions. In cases (ii) and (iii), we apply
optical tweezers incident along the $x$ direction on ions of index
\{..., 625, 635, 645, ...\} and create $\omega_{y}^{{\rm ot}}=0.2\omega_{x}$
($\sim2\pi\times1$ MHz) trapping frequency in addition along the
$y$ direction on these ions. (b) The required intensity of the laser
beam in terms of the maximal $\eta\Omega$ of the pulse shape in (a).
See text for the definitions for all quantities. \label{fig:infidelpower}}
\end{figure}
In this section, we demonstrate an elementary quantum gate, control
phase flip (CPF), on the $i$th and $j$th qubits by shining identical
laser beams on them of engineered pulse shapes. An ideal CPF is described
by $U_{ij}^{\rm{ideal}}=\exp\left(-{\rm i}\frac{\pi}{4}\sigma_{i}^{z}\sigma_{j}^{z}\right)$
with $\sigma_{i}^{z}$ the $z$ component of the Pauli matrices for
the $i$th qubit. Here we follow the pulse shaping scheme of quantum
gate design \cite{lin09,arbitraryspeed_zhu06} but the push gate protocols
\cite{microtraps_fastgates_hope18,fastgate_garciaripoll03,fastgate_duan04,fastgate_hope15}
may also apply without changing the main conclusion. This former scheme
is based on the Sorensen-Molmer-Milburn gate protocol, where a pair
of laser beams of slightly different colour are applied on the target
qubits, resulting in state-dependent momentum kicks parallel to the
wavevector difference $\Delta\mathbf{k}$ of beat note frequency $\mu$.
For quantum gates operating in the transverse (longitudinal) modes,
$\Delta\mathbf{k}$ needs to be along the $x$ or $y$ ($z$) direction.
In this work, we use transverse-mode gates for better protection from
the noise \cite{zhu06}. The parameter $\mu$ steers how motional
modes are involved for it amounts to modulating the Rabi frequency
$\Omega^{(i,j)}(t)=\Omega_{0}(t)\sin\mu t$, and hence the forces.
As a result, the gate dynamics can be given by the evolution operator
\begin{equation}
U(t) =\exp\left[\sum_{i,k}\left(\alpha_{i}^{k}(t)a_{k}^{\dagger}+\alpha_{i}^{k\ast}(t)a_{k}\right)\sigma_{i}^{z}+{\rm i}\sum_{i<j}\phi_{ij}(t)\sigma_{i}^{z}\sigma_{j}^{z}\right],\label{eq:gate_evolution}
\end{equation}
where $\alpha_{i}^{k}(\tau)=\int_{0}^{\tau}dt\Omega_{0}(t)\eta_{k}G_{ik}e^{{\rm i}\omega_{k}t}$
and $\phi_{ij}(\tau)=2\int_{0}^{\tau}dt^{\prime}\int_{0}^{t^{\prime}}dt\sum_{k}\eta_{k}^{2}G_{ik}G_{jk}\Omega_{0}(t)\Omega_{0}(t^{\prime})\\
\sin\omega_{k}(t^{\prime}-t)$
with the Lamb Dicke parameter $\eta_{k}=|\Delta\mathbf{k}|\sqrt{\hbar/(m\omega_{k})}$.
$G_{ik}$ is the element of the canonical transformation matrix that
diagonalizes $A$, i.e., $G^{\top}AG=A^{D}$ is diagonalised, corresponding
to the $i$th ion and the $k$th eigenmode. We shape the pulse profile
of gate duration $t_{g}$ by chopping it into $M$ equal time bins,
and for each bin the amplitude of the Rabi frequency is set differently,
i.e., $\Omega_{0}(t)=\Omega_{m}$ for $(m-1)t_{g}/M<t\le mt_{g}/M$.
The evolution operator (\ref{eq:gate_evolution}) leads to a gate
error, called the computational infidelity $\delta F_{{\rm c}}\equiv[6-2(\Gamma_{i}+\Gamma_{j})-\Gamma_{+}-\Gamma_{-}]/8$,
where $\Gamma_{i}=\exp[-\sum_{k}|\alpha_{i}^{k}(t_{g})|^{2}\bar{\beta}_{k}/2]$
and $\Gamma_{\pm}=\exp[-\sum_{k}|\alpha_{i}^{k}\pm\alpha_{j}^{k}|^{2}\bar{\beta}_{k}/2]$
with $\bar{\beta}_{k}=\coth[\hbar\omega_{k}/(k_{B}T)]$.

The local motion consideration suggests that we take into account
only a small subset of local ions. We thus model a quantum gate by
considering a cell from an infinitely long ion crystal and freezing
all the ions outside the cell. We then obtain the pulse shape from
optimisation, and apply these parameters to the actual configuration
of an ion crystal. For better estimation for the pulse shape, the
cell of simulation must have a larger size than the target gate distance
in order to take into account needed degrees of freedom. The more
degrees are included, the more accurate of the gate simulation, along
with the more computational load. For example, here we consider an
$N=1795$ ion crystal stabilised by a series of optical tweezers shining
on ions of index \{..., 625, 635, 645, ...\} along the $x$ direction.
To obtain a pulse shape that implements a quantum gate between the
$628$th and $631$st ions, we may turn to consider a gate between
the $3$rd and $6$th ions within a 9-ion cell. The infidelity $\delta F_{{\rm c}}$
for varied gate times are plotted in Fig.\,\ref{fig:infidelpower}(a).
For the $x$-mode gates, the infidelity on a large array appears almost
identical to the small array simulations, implying the validity of
the local mode consideration. Note that optical tweezers play no role
here since they have no effect on the motional spectrum of the $x$
direction. But for the $y$-mode gates, increase in infidelity can
be expected if optical tweezers alter the relevant mode spectrum.
For comparison, we also look at the case with gate operation between
the $624$th and $627$th ions (the $625$th is tweezered). We find
that the gate error increases significantly for $t_{g}$ shorter than
$2$ $\mu$s, as shown in Fig.\,\ref{fig:infidelpower}(a). This
is caused by intervention of the tweezered ion in between. For longer
gate times ($t_{g}\sim 2$ $\mu s$ and larger), surprisingly, the effect of
tweezer frequency vanishes. Since gates of longer gate times more
rely on interplay of the slower modes ($\sim 100$ kHz), insertion
of optical tweezers of a few megahertz plays little role here. Also,
in Fig.\,\ref{fig:infidelpower}(b), we show the required laser power
for various gate times in terms of the maximal Rabi frequency $\eta\Omega_{\max}$
of the pulse shape, and observe a power law $\eta\Omega_{\max}\sim t_{g}^{-1.53}$.
For gates longer than $20$ $\mu$s, the required $\eta\Omega_{\max}$
drops to about hundreds of kilohertz, a typical range of laser power
currently available.

In addition to the computational infidelity $\delta F_{{\rm c}}$
originating from imperfect disentangling between qubits and motion
at the computing stage, there are other major sources of error associated
with the temperature. The first one is owing to breakdown of Lamb-Dicke
approximation when the thermal fluctuations in ions' displacement
become non-negligible compared to the wavelength of the laser field.
This error of a transverse-mode gate has been shown to be $\delta F_{{\rm LD}}^{\xi=x,y}\approx\pi^{2}\eta_{\xi}^{4}(\bar{n}_{\xi}^{2}+\bar{n}_{\xi}+1/8)$
\cite{zhu06}, and $\bar{n}_{\xi}$ is the mean phonon number of the
$\xi$ mode. For large $\bar{n}_{\xi}$, this form can be recast as
$\delta F_{{\rm LD}}^{\xi}\approx\pi^{2}(\Delta k)^{4}(\delta x_{{\rm th}}^{\xi2})^{2}$,
where $\delta x_{{\rm th}}^{\xi}$ is the mean position fluctuation
of a motional thermal state. The second source of error comes from
the high-order contribution beyond the harmonic approximation for
normal modes. This error can be characterised by $\delta F_{{\rm a}}^{\xi=x,y,z}\approx(\delta x_{{\rm th}}^{\xi}/d_{0})^{2}$
\cite{lin09}. The third one is associated with the single qubit addressability.
As an ion is illuminated by a laser beam of Gaussian profile $\Omega\sim\exp[-(z-z_{0})^{2}/w^{2}]$,
where $z_{0}$ is the equilibrium position of the ion and $w$ is
the beam size, it sees spatially dependent field intensity due to
its longitudinal motion. This non-uniformity of the driving field
introduces an infidelity $\delta F_{{\rm b}}^{z}=(\pi^{2}/4)(\delta z_{{\rm th}}/w)^{4}$.
In the following section, we will see how these infidelities are taken
care of through cooling.

\section{Cooling\label{sec:Cooling}}

A large array of ions is subjected to serious heating and needs to
be continuously cooled not just for stabilising the array structure
but also sustaining the quantum coherence. When some ion qubits have
carried quantum information and cannot be directly cooled, one can
still cool their neighbours (coolant ions) to help remove the heat
through momentum exchange. Such a sympathetic cooling scheme has been
commonly used and studied \cite{symcool_kielpinsli00,symcool_barrett03,symcool_home09,symcool_brown11,lin_qip16}.
In this section, we discuss the performance of this scheme in the
presence of optical tweezers.

\begin{figure}[t]
\begin{raggedleft}
\includegraphics[width=11cm]{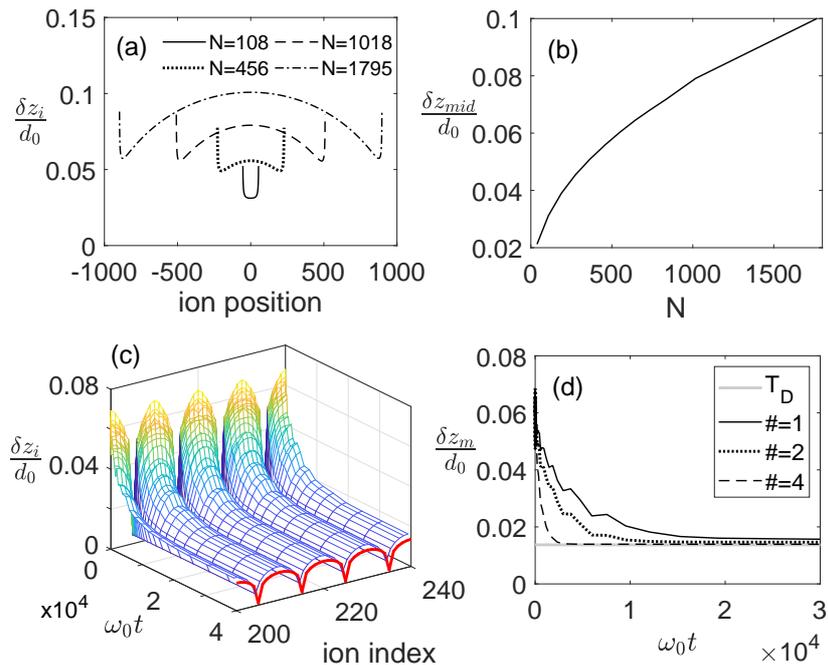}
\par\end{raggedleft}
\centering{}\caption{(a) The longitudinal position fluctuation (PF) of ion arrays of different
length without optical tweezers. (b) The longitudinal PF $\delta z_{{\rm mid}}$
corresponding to the middle of the ion crystal as a function of the
system size $N$ without optical tweezers. Note that the curve can
be fitted by $\delta z_{{\rm mid}}/d_{0}=4.3\times10^{-3}N^{0.42}$,
which reaches 1 $\mu$m when $N\sim1800$. (c) Cooling dynamics
in terms of $\delta z_{i}$ of the $200$th to $240$th ions on an
$N=1018$ ion crystal with optical tweezers and sympathetic cooling.
Here, optical tweezers are applied on ions of index $\{...,205,215,225,...\}$;
Doppler cooling beams are applied on ions of index $\{...,206,216,226,...\}$.
For reference, the red curve shows the PF profile at $T=T_{{\rm D}}$.
(d) The temporal relaxation of the longitudinal PF of the $240$th
ion, $\delta z_{{\rm m}}$, corresponding to the middle one within
the cell (locally maximally displaced). The solid curve labeled by
$\#=1$ corresponds to the case in (c), with a single Doppler cooling
beam for each cell. The dotted curve by $\#=2$ and the dashed curve
by $\#=4$ correspond to the cases with two and four Doppler beams
for each cell, respectively, applied on the adjacent ions of both
sides of each tweezered ion equally. By fitting to an exponential
profile ($\delta z_{{\rm m}}=ae^{-t/\tau_{R}^{z}}+\delta z_{{\rm s}}$),
we have the relaxation times $\tau_{{\rm R}}^{z}$ of $4900\omega_{0}^{-1}$,
$2590\omega_{0}^{-1}$, and $677\omega_{0}^{-1}$ and $\delta z_{{\rm s}}=0.016d_{0}$,
$0.015d_{0}$, and $0.014d_{0}$, respectively. The horizontal reference
line corresponds to the middle ion's longitudinal PF ($=0.014d_{0}$)
at $T=T_{{\rm D}}$.}
\label{fig:symcool}
\end{figure}

\begin{figure}[t]
\begin{raggedleft}
\includegraphics[width=15cm]{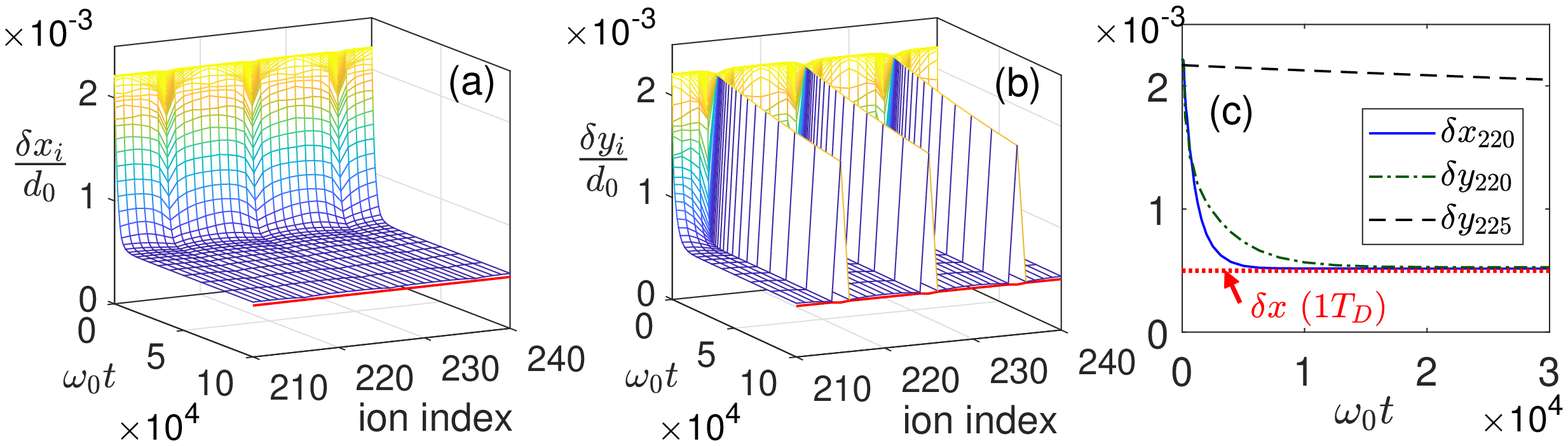}
\par\end{raggedleft}
\centering{}\caption{Transverse cooling dynamics in terms of (a) $\delta x_{i}$ and (b)
$\delta y_{i}$ for the $210$th to $240$th ions on an $N=1018$
ion crystal with optical tweezers (along the $x$ direction) and sympathetic
cooling on the $x$ and $y$ modes, respectively. The arrangement
of optical tweezers is the same as Fig.\,\ref{fig:symcool}(c). (c)
The temporal profiles of the transverse PFs of the $220$th ion, $\delta x_{{\rm m}}$
and $\delta y_{{\rm m}}$, corresponding to the middle one within
the cell. By fitting to an exponential profile ($\delta x_{{\rm m}}^{\xi}=ae^{-t/\tau_{{\rm R}}^{\xi}}+\delta x_{{\rm s}}^{\xi}$),
we have the relaxation times $\tau_{{\rm R}}^{x}=113\omega_{0}^{-1}$,
$\delta x_{{\rm s}}=5.1\times10^{-4}d_{0}$ for the $x$ modes, and
$\tau_{{\rm R}}^{y}=302\omega_{0}^{-1}$, $\delta y_{{\rm s}}=5.3\times10^{-4}d_{0}$
for the $y$ modes. For comparison, the profile of the $225$th (tweezered)
ion is plotted with $\tau_{{\rm R}}^{y}=1.25\times10^{4}\omega_{0}^{-1}$
and $\delta y_{{\rm s}}=1.6\times10^{-3}d_{0}$. The horizontal reference
line corresponds to the middle ion's PF ($=5.1\times10^{-4}d_{0}$)
at $T=T_{{\rm D}}$. \label{fig:symcool_trans}}
\end{figure}
The mathematical model for sympathetic cooling has been discussed
in \cite{lin_qip16}, where every ion is coupled to its own thermal
bath but linked with each other by Coulomb interaction. The Heisenberg-Langevin's
equation for this system is given by
\begin{equation}
\eqalign{
\dot{x}_{i}^{\xi} & =p_{i}^{\xi}\\
\dot{p}_{i}^{\xi} & =-\sum_{j}A_{ij}^{\xi}x_{j}^{\xi}-\gamma_{i}^{\xi}p_{i}^{\xi}+\sqrt{2\gamma_{i}^{\xi}}\eta_{i}^{\xi},}
\end{equation}
where $x_{i}^{\xi}$ and $p_{i}^{\xi}$ are the $i$th ion's coordinate
and momentum operators along the $\xi$ direction in units of $d_{0}$
and $m\omega_{0}d_{0}$, respectively, $\gamma_{i}^{\xi}$ is the
damping rate characterising the coupling between the ion and its environment,
and $\eta_{i}^{\xi}$ corresponds to random kicks owing to the baths.
For Markov baths, we take $\langle\eta_{i}^{\xi}(t)\eta_{j}^{\xi}(t^{\prime})\rangle=\delta_{ij}\delta(t-t^{\prime})\sum_{k}\tilde{\omega}_{k}G_{ik}^{2}(\bar{n}_{k}^{\xi,{\rm B}}(\tilde{T}_{i})+1/2)$,
where the mean phonon number of the thermal distribution $\bar{n}_{k}^{\xi,{\rm B}}(\tilde{T})=[\exp(\tilde{\omega}_{k}/\tilde{T})-1]^{-1}$
with $\tilde{\omega}_{k}=\omega_{k}/\omega_{0}$ and $\tilde{T}=k_{{\rm B}}T/(\hbar\omega_{0})$.
For coolant ions, we assume that their bath temperature is the Doppler
temperature $T_{{\rm D}}$, and the associated damping rate $\gamma_{i}=0.01$
corresponds to a typical retardation coefficient given by the Doppler
cooling. For rest of the ions, we model background heating by setting
$\gamma_{i}^{\xi}\tilde{T}_{i}=10^{-4}$, which amounts to a heating
rate of about 60 phonons per second per ion for the frequency level
$\sim\omega_{0}$. To study the worse case scenarios, we take $\tilde{T}_{i}\rightarrow\infty$
with $\gamma_{i}^{\xi}=10^{-4}/\tilde{T}_{i}$ and look at the resultant
position fluctuation (PF) defined by $\delta x_{i}^{\xi}=[\langle x_{i}^{2}\rangle-\langle x_{i}\rangle^{2}]^{1/2}$
of the $i$th ion.

Figure \ref{fig:symcool}(a) shows the distribution of the longitudinal
PF for various array sizes without optical tweezers. Neglecting those
ions on the array edges, the greatest one $\delta z_{{\rm mid}}$
occur at the centre of the array. We find the scaling relation $\delta z_{{\rm mid}}/d_{0}=4.3\times10^{-3}N^{0.42}$
at $T_{{\rm D}}$ as shown in Fig.\,\ref{fig:symcool}(b). Even when
the whole system is Doppler cooled, $\delta z_{{\rm mid}}$ becomes
larger than 10\% ($1$ $\mu$m) of the ion spacing (10 $\mu$m) for
$N\sim1800$ and larger. This not only results in very large $\delta F_{{\rm a}}^{z}$,
$\delta F_{{\rm b}}^{z}$ (approximately or larger than $10^{-2}$) but also makes the whole array
unstable. Now we turn to the optical tweezer assisted configuration.
For demonstration, we place optical tweezers on the ion array with
a spatial period $P=10$, and simultaneously Doppler cool the neighbouring
ions next to the tweezered ones. Fig.\,\ref{fig:symcool}(c) shows
the cooling dynamics of the array starting from $T=20T_{\rm{D}}$
at $t=0$, where we observe that the overall $\delta z_{i}$ goes
down almost exponentially, and asymptotically approaches to a steady-state
value. Specifically, we look at the middle ion ($220$th) within a
cell defined by two tweezered ones ($215$th and $225$th), and plot
its PF relaxation in Fig.\,\ref{fig:symcool}(d). From this curve,
we extract a relaxation time $\tau_{{\rm R}}^{z}$ by fitting to an
exponential profile. Our calculation shows that $\tau_{{\rm R}}^{z}$
is on the order of a few milliseconds if one ion is Doppler cooled
within this cell. Increasing the number of coolant ions shortens $\tau_{R}$
down to sub-milliseconds. It can be also shown that the steady state
is very close to the Doppler temperature with PF suppressed down to
about $2$\% of $d_{0}$, resulting in much smaller $\delta F_{{\rm a}}^{z}\sim4\times10^{-4}$
and $\delta F_{{\rm b}}^{z}\sim10^{-5}$.

On the other hand, since the transverse frequency is on the order
of megahertz, the associated PFs $\delta x_{i}$ or $\delta y_{i}$
are about one order of magnitude smaller than $\delta z_{i}$. Note
that only the $y$-mode fluctuations are affected by the optical tweezers
applied along the $x$ direction. These fluctuations contribute majorly
to the gate fidelity $\delta F_{{\rm LD}}^{x,y}$ and $\delta F_{{\rm a}}^{x,y}$.
Given $\delta x_{{\rm th}}^{\xi}/d_{0}=5.1\times10^{-4}$ at $T_{{\rm D}}$,
$\delta F_{{\rm LD}}^{x,y}=7.0\times10^{-4}$ and $\delta F_{{\rm a}}^{x,y}\sim10^{-7}$.
In Fig.\,\ref{fig:symcool_trans}, we show the relaxation dynamics
of $\delta x_{i}$ and $\delta y_{i}$ under the similar arrangement
of Fig.\,\ref{fig:symcool}(c). The steady-state PF is almost identical
to that given at $T_{{\rm D}}$, resulting in similar infidelities.
On the other hand, we find that the $x$-mode PF presents much faster
relaxation without effects of optical tweezers compared to the $y$-mode
PF. This contradicts to the intuition we hold for the longitudinal
mode, where the optical tweezers play a positive role that improves
cooling efficiency. Due to frequency mismatch of the local tweezers
to the neighbouring ions, the heat tends to be trapped on the tweezered
sites, exhibiting higher PFs and slower relaxation as shown in Fig.\,\ref{fig:symcool_trans}(b).

\section{Local trap and cooling\label{sec:Localtrap}}

\begin{figure}[t]
\begin{raggedleft}
\includegraphics[width=12cm]{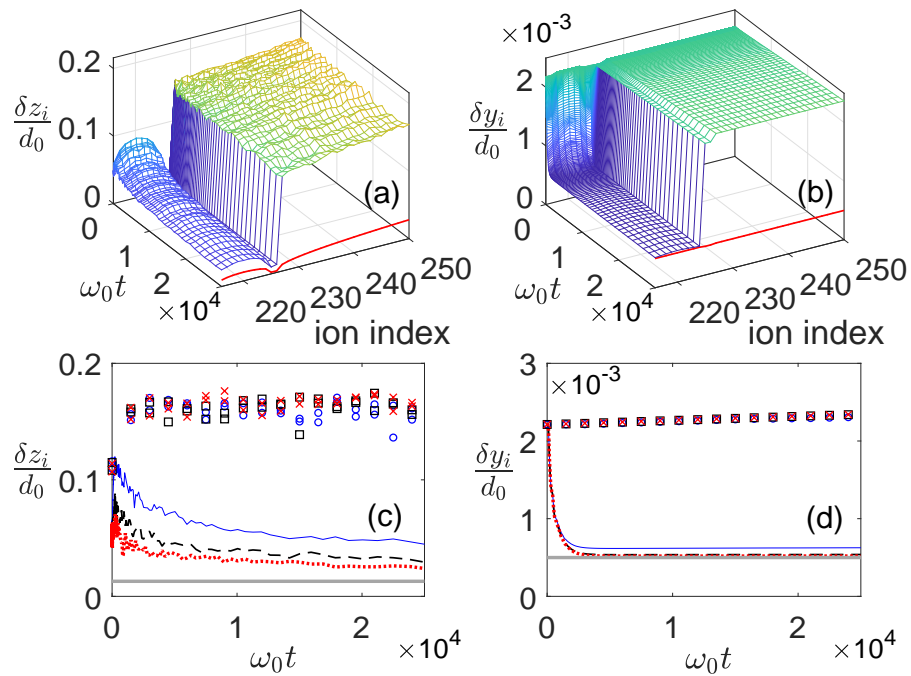}
\par\end{raggedleft}
\centering{}\caption{Cooling dynamics of (a) longitudinal and (b) transverse PFs, $\delta z_{i}$
and $\delta y_{i}$, respectively, on an $N=1018$ ion crystal. Optical
tweezers are applied on the ions of index $\{214,215,225,226\}$ while
the Doppler cooling is performed on the ions of index $\{216,217,223,224\}$.
For reference, the supposed PF profiles at the Doppler temperature
are plotted in red. We can see that the temperature in terms of PF
within the cell can be made much lower than that outside the cell
and to the Doppler one. Also, relaxation curves of the (c) longitudinal
and (d) transverse PFs, respectively, of the $220$th ion (the middle
one in the cell) are plotted. The solid (blue), dashed (black), and
dotted (red) curves correspond to cases with the walls of the cell
formed by one, two, and three tweezered ions, respectively, on each
side. When fitted to an exponential profile $\delta x^{\xi}=ae^{-t/\tau_{{\rm R}}^{\xi}}+\delta x_{{\rm s}}^{\xi}$,
we find $\tau_{{\rm R}}^{z}=3.9\times10^{4}\omega_{0}^{-1}$, $2.4\times10^{4}\omega_{0}^{-1}$,
$1.7\times10^{4}\omega_{0}^{-1}$ and $\delta z_{s}=0.031d_{0}$,
$0.025d_{0}$, $0.021d_{0}$, respectively, and $\tau_{{\rm R}}^{y}=658\omega_{0}^{-1}$,
$698\omega_{0}^{-1}$, $709\omega_{0}^{-1}$ and $\delta y_{{\rm s}}=6.1\times10^{-4}d_{0}$,
$5.3\times10^{-4}d_{0}$, $5.2\times10^{-4}d_{0}$, respectively.
For comparison, the horizontal reference lines correspond to the middle
ion's longitudinal and transverse PFs at $T=T_{{\rm D}}$. The PFs
corresponding to the $190$th and $250$th ions (30 sites away from
the cell center) are also plotted, showing the heating effect outside
the cell.\label{fig:localcool}}
\end{figure}
We have learned from previous analysis that optical tweezers effectively
play a role to serve as separators that partitions the whole ion crystal
into sub-arrays. A cell can be defined given two or two groups of
tweezered ions on both sides containing a sub-array. And the thickness
of a cell wall can be thought as the number of tweezered ions placed
next to each other. Intuitively speaking, a cell can be made less
coupled to the rest of the ion crystal in terms of momentum exchange
by increasing the wall thickness. Therefore, it is possible to cool
a cell that has its own temperature within a timescale of interest.
If this is the case, one can implement a faithful quantum gate within
a locally cooled cell without spending resources on cooling irrelevant
ions. In this section, we investigate the local dynamics of sympathetic
cooling on a chosen cell while other parts of the array may be gradually
heated by thermal noises.

Here we take an exemplary case by considering a large ion array of
$N=1018$, on which we choose a cell that contains ions of index $\{216,\cdots,224\}$.
Next to the edge ions, we shine optical tweezer beams to form two
walls of the same thickness for the cell. To sympathetically cool
those ions within, we apply Doppler cooling on the edge ions of index
$\{216,217,223,224\}$, and look at the local dynamics of PF. Fig.\,\ref{fig:localcool}(a)
and (b) show the profiles of $\delta z_{i}$ and $\delta y_{i}$,
respectively, of the $216\rm{th}\sim250$th ions, where we set the
initial temperature to be $20T_{{\rm D}}$. As expected, we find that
PFs within the cell decrease with time nearly exponentially and approach
to a steady-state distribution, which is found to be slightly larger
than that at $T_{{\rm D}}$ due to the presence of background heating.
In the case where the wall thickness is two on both sides, we have
the cooled (steady-state) PF $\delta z_{{\rm s}}\approx0.025d_{0}$
with relaxation time $\tau_{{\rm R}}^{z}\approx26$ ms, and $\delta y_{{\rm s}}\approx5.3\times10^{-4}d_{0}$
with $\tau_{{\rm R}}^{y}\approx0.78$ ms for the longitudinal and
transverse modes, respectively, thus keeping all the largest gate
infidelities discussed in the previous sections on the order of $10^{-4}$.

The steady-state fluctuation profile of the cell can be made arbitrarily
close to the ideally Doppler cooled case by increasing the number
of coolant ions and thickening the cell walls, as shown in Fig.\,\ref{fig:localcool}(c)
and (d) for longitudinal and transverse directions, respectively.
Heating can be generally observed for ions outside the cell. For the
longitudinal motion, the fluctuations appear to be slightly suppressed
for ions closer to the cell. But for the transverse one, the fluctuation
profile tends to be more independent of locations. This is because
the longitudinal modes are soft ones with longer oscillation periods
($\sim\omega_{{\rm L}}^{-1}$), allowing momentum exchange (characterised
by $\omega_{0}^{-1}$) to take place between distant ions over the
walls. This is generally not the case for transverse modes ($\sim\omega_{x,y}^{-1}$),
whose periods are too short to support exchange momentum between distant
ones.

\section{Conclusion\label{sec:conclusion}}

To sum up, we have proposed an ion crystal architecture stabilised
by optical tweezers. In a traditional RF ion trap, a large array usually
has collective modes of vanishing frequencies in the longitudinal
direction and hence divergent motional excitation given any finite
temperature, making its structure vulnerable. Though typically the
Coulomb interaction is much stronger than dipole forces given by optical
tweezers, the relevant motional frequency scale is determined by the
next-order contribution, the residual Coulomb interaction, characterised
by $\omega_{0}$, which is about hundreds of kilohertz for ion separation
about $10$ $\mu$m. Thus, application of optical tweezers is able
to alter the longitudinal motional spectrum and lifts the frequencies
of the ground modes by appropriately arranging the optical tweezers.
Further, it has been shown that ions illuminated by optical tweezers
can be seen effectively pinned in space by comparing the motional
spectra. Cells can be thus defined by being sandwiched by tweezered
ions, introducing the convenience to look at localised motion associated
with these cells.

For a large-scale computing, making operations depending on local
degrees of freedom is very important. This allows parallel processing
and pipelined task management. We have also shown faithful quantum
logic gates based on transverse modes for two free (not tweezered)
ions within a cell. It should be emphasized that these gates are implemented
using the controlling parameters obtained in a small array simulation.
This implies that these parameters have translation symmetry along
the ion crystal, that is, universal to anywhere of implementation
given the same gate distance. The presence of optical tweezers, depending
on the incident direction, may alter the motional spectrum that transverse
gates are based on. We have shown that the gate error can still be
kept lower than $10^{-5}$ unless optical tweezers intervene the ions
between the target qubits.

In addition to the computational error associated with the gate design,
we have also investigated other major sources of imperfection, such
as breakdown of the Lamb-Dicke approximation, anharmonicity, and a
non-uniform beam profile, which require cooling to reduce the longitudinal
and transverse PFs for insuring the gate fidelity. For this purpose,
we have examined the sympathetic cooling dynamics in the presence
of optical tweezers. The advantage of applying optical tweezers is
apparent: with regular arrangement of tweezers along the system, the
overall PFs can be greatly suppressed compared to the cases without
them, where the PFs basically diverge as the system size increases.
We can thus maintain the whole system close to the Doppler temperature
with thermal effects contributing only about $10^{-4}$ of gate errors.
Finally, we have looked into the locality of motion for a cell by
adding more tweeezered ions to increase the wall thickness. Note that
such locality is a remarkable feature because trapped ions are usually
collectively coupled so that, without optical tweezers, local manipulations
tend to cause long-range interferences. For the concern of parallel
computing, these interferences are unwanted effects but can now be
eliminated with the help of optical tweezers. We have shown that a
cell can be cooled almost independently from other parts of the system,
thus suggesting a more economical way to spending resources for cooling
issues. Further, we expect that our proposed scheme also interests
communities of quantum simulation of cavity quantum electrodynamics
for the resemblance between a cell defined by tweezers for phonons
and a real cavity for photons, and quantum thermodynamics for the
analogue of heat transfer.

\ack{}{We thank the support from MOST of Taiwan under Grant No. 105-2112-M-002-015-MY3
and National Taiwan University under Grant No. NTU-106R891708. GDL
thanks M.-S. Chang for valuable discussion and feedback.}

\end{document}